\documentclass[twocolumn,showpacs,preprintnumbers,amsmath,amssymb]{revtex4}

\usepackage{graphicx}
\usepackage{dcolumn}
\usepackage{bm}

\newcommand{\be}{\begin{equation}}
\newcommand{\ee}{\end{equation}}
\newcommand{\bdis}{\begin{displaymath}}
\newcommand{\edis}{\end{displaymath}}
\newcommand{\pd}{\partial}

\newcommand{\mbb}{\mathbb}

\begin{document}

\title{A short note on Casimir force and radius stabilization in QFT with non-commutative target space}

\author{Michal Demetrian}\email{demetrian@fmph.uniba.sk}
\affiliation{Comenius University \\
Mlynska Dolina F2, 842 48, Bratislava, Slovak Republic }

\date{\today}

\begin{abstract}
Stable radius of cylindrical space due to additional repulsion caused by noncommutativity of two-component field values is found. 
\end{abstract}

\maketitle

\section{Introduction}

Since the famous paper of Snyder \cite{snyder} the idea to build up quantum theory on noncommutative spaces has been widely studied by many
authors. Quantum physics on noncommutative space became one of the main trends in modern physics as possible tool to study how space-time structure itself acts upon matter at very small scales and how space-time itself can be understood as quantum object, or quantized. Appearance of Casimir force in quantum fields under some nontrivial geometrical or topological situations has been studied widely in both commutative and non-commutative cases, as for example in \cite{nam}. Special attention is often paid to the effect of radius stabilization of suitable manifolds. In this short note we transfer the notion of non-commutativity into the abstract space of values of quantized field in very simple case of scalar field. The idea of this work follows clearly explained treatment of the work \cite{bal1}, where the reader can find references to other related papers. 

\section{The commutative model}

Let us start with two (in fact, there is no way how to proceed with single-valued fields, on the other way, one may try to use the quaternions, see \cite{adler}) component real scalar $(\varphi^1,\varphi^2),\ \varphi^a=\varphi^a(x,t)$ defined over two dimensional
cylinder, with spatial period equal to $2\pi R$. This means we assume the periodic conditions:
$\varphi(x,t)=\varphi(x+2\pi R,t),\ \partial_x(x,t)=\partial_x(x+2\pi R,t)$ hold. Such fields can be decomposed into its Fourier spatial components as follows
\bdis
\varphi^a(x,t)=\sum_{n\in\mbb{Z}} e^{\frac{2\pi i}{R}nx}\phi_n^a(t), \quad \phi^a_n=\phi^{a\star}_{-n}, a=1,2,
\edis
where the last condition guarantees reality of $\varphi^a$.
Let us consider $\varphi^a$ be a pair of massless free scalars, this means
the dynamics is given by the lagrangian
\begin{eqnarray}
& &
L=\frac{1}{2}\int{\rm d}x\left[(\pd_t\varphi^1)^2+(\pd_t\varphi^2)^2-(\pd_x\varphi^1)^2-(\pd_x\varphi^2)^2\right] \nonumber \\
& &
=\frac{R}{2}\sum_{a,n}\left[ \dot{\phi}^a_n\dot{\phi}^a_{-n}-\left(\frac{2\pi n}{R}\right)^2\phi^a_n\phi^a_{-n}\right].
\end{eqnarray}
Defining momenta:
$$\pi^a_n=\frac{\pd L}{\pd\dot{\phi_n^a}}=R\dot{\phi}_{-n}^a=(\pi^a_{-n})^\star$$
one easily constructs the Hamiltonian of our system:
\be
H=\sum_{a}\frac{(\pi_0^a)^2}{2R}+\frac{1}{2R}\sum_{a,n\not=0}\left[\pi^a_n\pi^a_{-n}+(2\pi n)^2\phi^a_n\phi^a_{-n}\right] ,
\ee
where
\bdis
[\phi^a_m,\phi^b_n]=[\pi^a_m,\pi^b_n]=0,\ [\phi^a_m,\pi^b_n]=i\delta_{mn}\delta^{ab} .
\edis

\section{The model with noncommutative plane as target space}

Motivated by the work \cite{bal1} we shall consider the following modification of the canonical commutation relations
\begin{eqnarray} \label{ccr1}
\left[ \hat{\phi}^a_n,\hat{\phi}^b_m \right] & = & \frac{i}{R}\epsilon^{ab}\theta(n)\delta_{m+n,0} , \nonumber \\
\left[ \hat{\pi}^a_n,\hat{\pi}^b_m\right] & = & 0 , \\
\left[ \hat{\phi}^a_n,\hat{\pi}^b_m \right] & = & i\delta^{ab}\delta_{mn} , \nonumber
\end{eqnarray}
where the smearing function $\theta$ depends upon index of the mode:
\be \label{smf}
\theta(n)=\vartheta e^{-\frac{2\pi^2\sigma^2}{R^2}n^2} .
\ee
In this expression $\vartheta$ is the parameter of non-commutativity of dimension $L$. The commutation relations (\ref{ccr1}) (the first one considering
coordinates) can be rewritten for
spatially dependent field coordinates as the following equal-time commutation relation with Gaussian smearing on the right-hand side:
\bdis
\left[\hat{\varphi}^a(x,t),\hat{\varphi}^b(y,t)\right]=\frac{i\epsilon^{ab}\vartheta}{\sqrt{2\pi}\sigma}e^{-\frac{(x-y)^2}{2\sigma^2}} .
\edis
This reduces, in the limit $\sigma\to 0$, to the local commutation rule:
\bdis
\left[\hat{\varphi}^a(x,t),\hat{\varphi}^b(y,t)\right]=i\epsilon^{ab}\vartheta \delta(x-y) , 
\edis 
by means of the standard gaussian approximation of Dirac mapping. 
We shall consider, accordingly with \cite{bal1}, the system which dynamics is given by the following hamiltonian:
\be \label{hamnc}
H=\frac{1}{2gR}\sum_{a,n}\left[\hat{\pi}^a_n\hat{\pi}^a_{-n}+(2\pi n)^2\hat{\phi}^a_n\hat{\phi}^a_{-n} \right].
\ee
Now, the idea is, that the commutation relations (\ref{ccr1}) can be transformed to the canonical ones by the linear transformation \cite{jade}:
\begin{eqnarray*}
\hat{\phi}^a_n & = & \phi^a_n-\frac{1}{2R}\epsilon^{ab}\theta(n)\pi^b_{-n} , \\
\hat{\pi}^a_n & = & \pi^a_{n} .
\end{eqnarray*}
This transforms our hamiltonian (\ref{hamnc}) into the form:
\begin{eqnarray}  \label{hamnc2}
& &
H=\sum_{a,n}\left\{\left[ 1+\left(\frac{\pi n\theta(n)}{R}\right)^2\pi^a_n\pi^a_{-n}\right]+(2\pi n)^2\phi^a_n\phi^a_{-n}\right\} \nonumber \\
& &
-\sum_{a,b,n}\frac{(2\pi n)^2}{R}\epsilon^{ab}\theta(n)\phi^a_n\pi^b_n .
\end{eqnarray}
The second term in this hamiltonian is proportional to
the $(1,2)$ component of the "angular momentum"\ operator corresponding to the rotations in field plane. This term can be interpreted as an
effective interaction in field plane with auxiliary magnetic-like field.
The hamiltonian can be simply diagonalised with help of properly chosen annihilation and
creation operators. In order to do this, let us define some useful notations:
\bdis
\omega_n=\frac{2\pi |n|}{R}, \ \Omega_n^2=1+\left(\frac{\pi n\theta(n)}{R}\right)^2 .
\edis
We introduce the operators:
\bdis
a_n^a=\sqrt{\frac{\Delta_n}{2}}\left(\phi^a_n+i\frac{\pi^a_{-n}}{\Delta_n}\right), \
a_n^{a\dag}=\sqrt{\frac{\Delta_n}{2}}\left(\phi^a_{-n}-i\frac{\pi^a_{n}}{\Delta_n}\right),
\edis
where
\bdis
\Delta_n=\frac{R\omega_n}{\Omega_n^2}=\frac{2\pi |n|}{\Omega_n^2} .
\edis
This pair of operators obeys the canonical commutation relations:
\bdis
\left[ a^a_n,a^b_m\right]=\left[ a^{a\dag}_n,a^{b\dag}_m\right]=0,\ \left[ a^a_n,a^{b\dag}_m\right]=\delta_{mn}^{ab},
\edis
and our hamiltonian (\ref{hamnc2}) is given by
\begin{eqnarray} \label{hamnc3}
& &
H=\sum_{a}\frac{(\pi_0^a)^2}{2R}+ \nonumber \\
& &
\sum_{a,n\not=0}\omega_n\Omega_n^2a_n^{a\dag}a_n^a+i\frac{1}{2}\sum_{a,b,n\not=0}\theta(n)\omega_n^2\epsilon^{ab}a_n^{a\dag}a_n^b .
\end{eqnarray}
The following canonical transformation
\bdis
A_n^1=\frac{1}{\sqrt{2}}\left( a_n^1-ia_n^2\right),\ A_n^2=\frac{1}{\sqrt{2}}\left( a_n^1+ia_n^2\right)
\edis
transforms our hamiltonian into the final diagonal form:
\be \label{hamnc4}
H=\sum_{n\not=0}\omega_n\left[ \Lambda_nA_n^{1}A_n^{1\dag}+\Lambda_n^2A_n^{2}A_n^{2\dag}\right] ,
\ee
where we have introduced the notation:
\bdis
\Lambda_n^1=\Omega_n^2-\frac{\pi |n|\theta(n)}{R},\ \Lambda_n^2=\Omega_n^2+\frac{\pi |n|\theta(n)}{R} .
\edis
This result tells us that the magnetic like term causes splitting of the energy levels of the free-field hamiltonian.

\section{Casimir energy}
We shall rewrite the hamiltonian (\ref{hamnc4}) into a more suitable form:
\begin{eqnarray} \label{hamnc5}
& &
H=\sum_{n\not=0}\left\{\omega_n\Omega_n^2\left[ A_n^{1}A_n^{1\dag}+A_n^{2}A_n^{2\dag}\right]\right.+ \nonumber \\
& &
\left. \frac{\pi |n|\theta(n)}{R}\left[ A_n^{2}A_n^{2\dag}-A_n^{1}A_n^{1\dag}\right]\right\}.
\end{eqnarray}
We are interested in vacuum expectation value of this operator. It is evident that the $1\leftrightarrow 2$-antisymmetric term does not
contributes to the vacuum expectation value. We can write
\begin{eqnarray*}
& &
E_C=\langle 0|H|0\rangle= \\
& &
\sum_{n\not=0}\omega_n\Omega_n^2\left\langle 0\left[ A_n^{1}A_n^{1\dag}+A_n^{2}A_n^{2\dag}\right] 0\right\rangle = \\
& &
\sum_{n\not=0}\omega_n\left\langle 0\left[ A_n^{1}A_n^{1\dag}+A_n^{2}A_n^{2\dag}\right] 0\right\rangle+ \\
& &
\frac{\pi^2 g^2}{R^2}\sum_{n\not=0}\omega_n n^2\theta^2(n)\left\langle 0\left[ A_n^{1}A_n^{1\dag}+A_n^{2}A_n^{2\dag}\right] 0\right\rangle\equiv \\
& &
E_C^0+E_C^1 .
\end{eqnarray*}
The term $E_C^0$ that does not contain noncommutativity parameter is the standard one and its value (after necessary regularization) can be find
in e.g. \cite{birelldavis}, or in an exhaustive explanation of methods of computation is given in the paper \cite{mostepanenko}. $E_C^0$ reads
\bdis
E_C^{0ren}=-\frac{1}{6 R} .
\edis
The $\vartheta$-dependent contribution to the Casimir energy is already regularized by $\sigma$ in (\ref{smf}) and reads
\bdis
E_C^1(\sigma)=\frac{4\pi^3\vartheta^2}{R^3}\sum_{n=1}^\infty n^3e^{-\frac{2\pi^2\sigma^2}{R^2}n^2} .
\edis
Renormalizing it we obtain
\bdis
E_C^{1ren}=\frac{4\pi^3\vartheta^2}{R^3}\zeta(-3)=\frac{\pi^3\vartheta^2}{30R^3} .
\edis
Finally, we have the Casimir energy given by
\be \label{ce}
E_C=-\frac{1}{6 R}+\frac{\pi^3\vartheta^2}{30R^3} .
\ee

\section{Discussion}

If $\vartheta\not=0$ then the Casimir energy (\ref{ce}) is not a monotonous function of the radius $R$. This fact makes the formula for the Casimir
energy in our situation different from the standard result that is obtained letting $\vartheta=0$. There is a radius $R_0$ at which the Casimir energy
attains its minimal value, namely
\be \label{extrem}
R_0=\left(\frac{3\pi^3}{5}\right)^{1/2}\vartheta .
\ee
We see that the non-locality of the field field commutators
generates in (\ref{ce}) the repulsive force at small distances $R<R_0$, and effectively this force can stabilize the radius of the space.

\acknowledgments{This work was supported by the grant scheme VEGA 1/0785/19 of the Ministry of Education, Science, and Sport of the Slovak Republic.}

\end{document}